\documentclass[%
 reprint,
superscriptaddress,
 amsmath,amssymb,
 aps,
]{revtex4-2}

\usepackage{graphicx}
\usepackage{dcolumn}
\usepackage{bm}
\usepackage{xcolor}
\usepackage{physics}

\begin{document}

\preprint{APS/123-QED}

\title{Floquet engineering in the presence of optically excited carriers}

\author{Mitchell A.\ Conway}
\affiliation{Optical Sciences Centre, Swinburne University of Technology, Hawthorn, 3122, Victoria, Australia}
\affiliation{ARC Centre of Excellence in Future Low-Energy Electronics Technologies, Swinburne University of Technology, Hawthorn, 3122, Victoria, Australia}

\author{Jonathan O.\ Tollerud}
\affiliation{Optical Sciences Centre, Swinburne University of Technology, Hawthorn, 3122, Victoria, Australia}
\affiliation{ARC Centre of Excellence in Future Low-Energy Electronics Technologies, Swinburne University of Technology, Hawthorn, 3122, Victoria, Australia}

\author{Thi-Hai-Yen Vu}
\affiliation{ARC Centre of Excellence in Future Low-Energy Electronics Technology, Monash University, Clayton, 3800, Victoria, Australia}
\affiliation{School of Physics and Astronomy, Monash University, Clayton, 3800, Victoria, Australia}

\author{Kenji Watanabe}
\affiliation{Research Center for Electronic and Optical Materials, National Institute for Materials Science, 1-1 Namiki, Tsukuba 305-0044, Japan}

\author{Takashi Taniguchi}
\affiliation{Research Center for Materials Nanoarchitectonics, National Institute for Materials Science,  1-1 Namiki, Tsukuba 305-0044, Japan}

\author{Michael S.\ Fuhrer}
\affiliation{ARC Centre of Excellence in Future Low-Energy Electronics Technology, Monash University, Clayton, 3800, Victoria, Australia}
\affiliation{School of Physics and Astronomy, Monash University, Clayton, 3800, Victoria, Australia}

\author{Mark T.\ Edmonds}
\affiliation{ARC Centre of Excellence in Future Low-Energy Electronics Technology, Monash University, Clayton, 3800, Victoria, Australia}
\affiliation{School of Physics and Astronomy, Monash University, Clayton, 3800, Victoria, Australia}
\affiliation{ANFF-VIC Technology Fellow, Melbourne Centre for Nanofabrication, Victorian Node of the Australian National Fabrication Facility, Clayton, VIC 3168, Australia}

\author{Jeffrey A.\ Davis}
\affiliation{Optical Sciences Centre, Swinburne University of Technology, Hawthorn, 3122, Victoria, Australia}
\affiliation{ARC Centre of Excellence in Future Low-Energy Electronics Technologies, Swinburne University of Technology, Hawthorn, 3122, Victoria, Australia}
\email{JDavis@swin.edu.au}

\date{\today}

\begin{abstract}
Floquet engineering provides an optical means to manipulate electronic bandstructures, however, carriers excited by the pump field can lead to an effective heating, and can obscure measurement of the band changes. A recent demonstration of the effects of Floquet engineering on a coherent ensemble of excitons in monolayer WS$_2$ proved particularly sensitive to non-adiabatic effects, while still being able to accurately resolve bandstructure changes. 
Here, we drive an AC-Stark effect in monolayer WS$_2$ using pulses with constant fluence but varying pulse duration (from 25-235~fs). With shorter pump pulses, the corresponding increase in peak intensity introduces additional carriers via two-photon absorption, leading to additional decoherence and peak broadening (which makes it difficult to resolve the AC-Stark shift). We use multidimensional coherent spectroscopy to create a coherent ensemble of excitons in monolayer WS$_2$ and measure the evolution of the coherence throughout the duration of the Floquet pump pulse. Changes to the amplitude of the macroscopic coherence quantifies the additional broadening. At the same time, the evolution of the average phase allows the instantaneous changes to the bandstructure to be quantified, and is not impacted by the additional broadening. 
This approach to measuring the evolution of Floquet-Bloch states demonstrates a means to quantify effective heating and non-adiabaticity caused by excited carriers, while at the same time resolving the coherent evolution of the bandstructure.
\end{abstract}


\maketitle

\section{\label{sec:intro}Introduction}

Floquet theory \cite{shirley1965solution} describes the modification of state energies, and in condensed matter the bandstructure, by a periodic perturbation.
Coherent light-matter interactions, for example, lead to the formation of photon-dressed band replicas (Floquet-Bloch bands) \cite{wang2013observation,zhou2023floquet}, and can enable transient control of bandgap energies via an AC-Stark effect \cite{kim2014ultrafast,sie2015valley} or produce exotic non-equilibrium electronic states \cite{kobayashi2023floquet,Lindner2011}. For example, interacting Floquet-Bloch states have been demonstrated to facilitate a topological phase transition in graphene \cite{mciver2020light}. Similar effects have been predicted to arise in monolayer transition metal dichalcogenides (TMDCs) \cite{sie2015valley,claassen2016all}, however, observations of a topological phase transition in these systems remain elusive. This is partially due to the challenges associated with disentangling light-induced bandstructure changes from photo-excited population effects. Furthermore, understanding the role of excited carriers and the resultant heating in the Floquet bandstructure is a key challenge to realising practical applications of Floquet engineering \cite{seetharam2015controlled,rudner2020band,aeschlimann2021survival}.

\begin{figure*}
\includegraphics{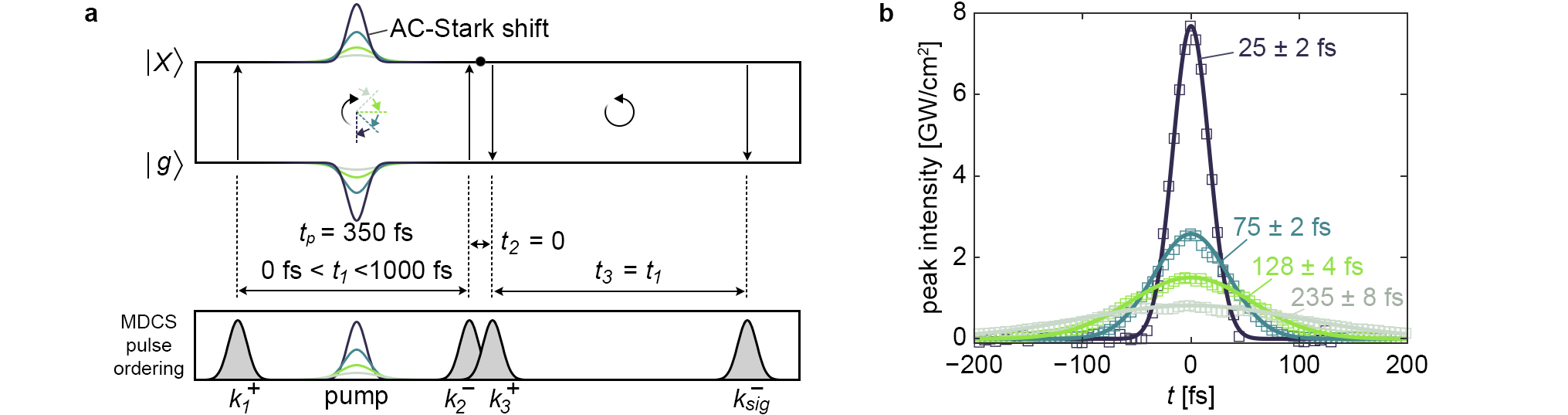}
\caption{\label{fig1}(a) Schematic representation of our pumped 1Q-MDCS measurements. The arrival of the first excitation pulse $\textbf{k}_1^{\ +}$ creates a coherent superposition between the ground state $\ket{g}$ and the $K^\prime$ valley exciton state, $\ket{X}$. This 1Q coherence evolves during the first inter-pulse delay $t_1$. With the delay between $\textbf{k}_2^{\ -}$ and $\textbf{k}_3^{\ +}$, $t_2 = 0$, the coherence $\ket{g}\bra{X}$ evolves until the simultaneous arrival of pulses $\textbf{k}_2^{\ -}$ and $\textbf{k}_3^{\ +}$. After the three excitation pulses interact with the sample, a third-order polarization is generated which radiates as the four-wave mixing signal, with wavevector $\textbf{k}_{sig}^{\ -} = -\textbf{k}_1^{\ +} + \textbf{k}_2^{\ -} + \textbf{k}_3^{\ +}$. Following the same approach used in Ref.\ \cite{conway2023effects}, we introduce a red-detuned $\sigma^+$ polarized pump pulse arriving $\sim$350~fs before $\textbf{k}_2^{\ -}$ and $\textbf{k}_3^{\ +}$, indicated by the colored Gaussian pulses. As a result, the phase of the $\ket{g}\bra{X}$ coherence oscillates faster while the pump is present due to an AC-Stark shift of the $\ket{g}$ and $\ket{X}$ state energies, which induces additional phase rotation of the measured signal. (b) Pump pulse shapes acquired from the deconvolution of a cross-correlation between the pump and $\textbf{k}_1^{\ +}$ beam. Throughout this work, we utilize four different pump beams which are 25, 75, 128, and 235~fs in duration with corresponding peak intensities of 7.7, 2.6, 1.5, and 0.8~GW/cm$^2$, respectively. All pump beams have the same total fluence of 205~$\mu$J/cm$^2$, and are centered at 1.82~eV (redshifted from the A-exciton resonance by $\sim$260~meV).}
\end{figure*}

There have been numerous reports of monolayer TMDCs driven by a red-detuned Floquet pump pulse that use transient absorption measurements to show changes to the bandgap induced by the AC-Stark effect \cite{kim2014ultrafast,sie2015valley,sie2017blochsiegert,ye2017optical,lamountain2018valley,yong2018biexcitonic,cunningham2019resonant,morrow2020quantum,kobayashi2023floquet,Earl2021}. Recently, the effects of Floquet engineering on a coherent ensemble of excitons was investigated in monolayer tungsten disulphide (WS$_2$) using multidimensional coherent spectroscopy (MDCS) \cite{conway2023effects}. The coherent ensemble proved particularly sensitive to non-adiabatic effects, revealing an effective power broadening phenomenon at high pump fluences. This broadening occurred only during the pump pulse, and was attributed to interactions between the virtual excitations created by the below resonance Floquet driving field and the excitons in the coherent ensemble. Despite these interactions and resultant broadening, the heterodyne detected signal phase smoothly followed the dynamics of the pump envelope, suggesting the average signal phase measured is not impacted by losses in the macroscopic coherence of the ensemble. This picture is consistent with previous theoretical work demonstrating that Floquet effects are present despite dissipation \cite{sato2019microscopic}, which suitably described a Floquet-topological phase transition in graphene \cite{mciver2020light}. Here we demonstrate that MDCS can reliably separate changes to the bandstructure from broadening caused by excited carriers; an important demonstration in the pursuit of realizing a Floquet-topological phase transition in monolayer TMDCs.

Here we perform MDCS measurements with an additional red-detuned pump pulse to drive an AC-Stark shift, but vary the pump duration, while keeping the total fluence the same. We find that as the duration of the pump pulses is decreased the corresponding increase in the peak intensity introduces additional carriers via two-photon absorption. In this regime, Floquet engineering effects are convolved with incoherent carrier injection which acts to obscure signatures of the bandstructure changes. In particular, these additional carriers lead to increased scattering and decoherence, which we are able to quantify. However, despite these losses in macroscopic coherence, we find that the evolution of the phase of the MDCS signal, which follows the average phase of the excitonic coherent superposition, is not impacted. From this phase evolution we are able to extract changes to the bandstructure, even in the presence of substantial photo-excited populations.

\section{\label{sec:methods}Experimental Methods}

In our MDCS measurements, three light-matter interactions trigger a third-order sample response, which is measured as a function of the inter-pulse delays \cite{tollerud2017coherent}. In one-quantum MDCS measurements (1Q-MDCS), the arrival of the first excitation pulse $\textbf{k}_1^{\ +}$ (with wavevector $\textbf{k}_1$ and the `+' indicating $\sigma^+$ circular polarization) creates a coherent superposition between the ground state and the $K^\prime$ exciton state, $\ket{g}\bra{X_{K^\prime}}$. This 1Q coherence evolves during the first inter-pulse delay, $t_1$, where the evolution of $\ket{g}\bra{X_{K^\prime}}$ is proportional to exp$[(-i\omega_{K^\prime}t_1)]$, and is measured by scanning the arrival time of $\textbf{k}_1^{\ +}$. The simultaneous arrival of the final two excitation pulses, with wavevectors $\textbf{k}_2$ and $\textbf{k}_3$ and circular polarization $\sigma^-$ and $\sigma^+$, respectively, generates a third-order polarization ($\ket{X_K}\bra{g}$) which radiates as the four-wave mixing signal in the direction given by $\textbf{k}_{sig}^{\ -}=-\textbf{k}_1^{\ +} + \textbf{k}_2^{\ -} + \textbf{k}_3^{\ +}$. The four-wave mixing signal is overlapped with a reference pulse ($\textbf{k}_{LO}^{\ -}$) for heterodyne detection, and spectrally resolved to give the amplitude and phase of the signal as a function of the emission energy, $\hbar\omega_3$. A schematic representation of the MDCS pulse ordering in our measurements is depicted in Fig.~\ref{fig1}(a). The MDCS excitation pulses were 30~$\pm$~1~fs in duration, with a fluence of 1.2~$\pm$~0.2~$\mu$J/cm$^2$, and focused to a 36~$\pm$~5~$\mu$m spot at the sample surface (see Supplementary Material for the spectrum of the MDCS pulses).

To measure an AC-Stark effect in monolayer WS$_2$, we employed the same technique outlined in Ref.\ \cite{conway2023effects}. The $\sigma^+$ polarized pump was red-detuned by $\sim$260 meV from the low temperature A-exciton energy of 2.087~eV, and set to arrive 350~fs before $\textbf{k}_2^{\ -}$ and $\textbf{k}_3^{\ +}$. To control the pump duration, we utilized a pulse shaper based around a spatial-light modulator \cite{vaughan2005diffraction}, which enabled control of the spectral phase and amplitude of our pulses. The pulse shaper was first used to ensure a flat spectral phase across the pump pulse using the multiphoton intra-pulse interference phase scan technique \cite{xu2006quantitative}. This yielded a 25~$\pm$~2~fs duration for the full spectral bandwidth of the pump pulse output from a non-colinear optical parametric amplifier (see Supplementary Material for pump spectrum). To increase the pulse duration, the spectral bandwidth was reduced by the pulse shaper, while maintaining a flat spectral phase. This has the consequence of reducing the pump fluence, so to ensure the total integrated energy of the pump pulse was constant, a variable neutral density filter was used to control the intensity. The measurements for all pulse durations were conducted with pump fluence of 205~$\pm$~10~$\mu$J/cm$^2$ with a pump spot size of 67~$\pm$~5~$\mu$m at the sample surface. Figure~\ref{fig1}(b) shows the temporal profile for each of the pump pulses utilized in this work, measured by deconvolving the cross-correlation of the pump and the $\textbf{k}_1^{\ +}$ excitation beam.

\section{Results and Discussion}

\subsection{Quantifying loss of macroscopic coherence due to two-photon absorption}

\begin{figure}
\includegraphics[width=\linewidth]{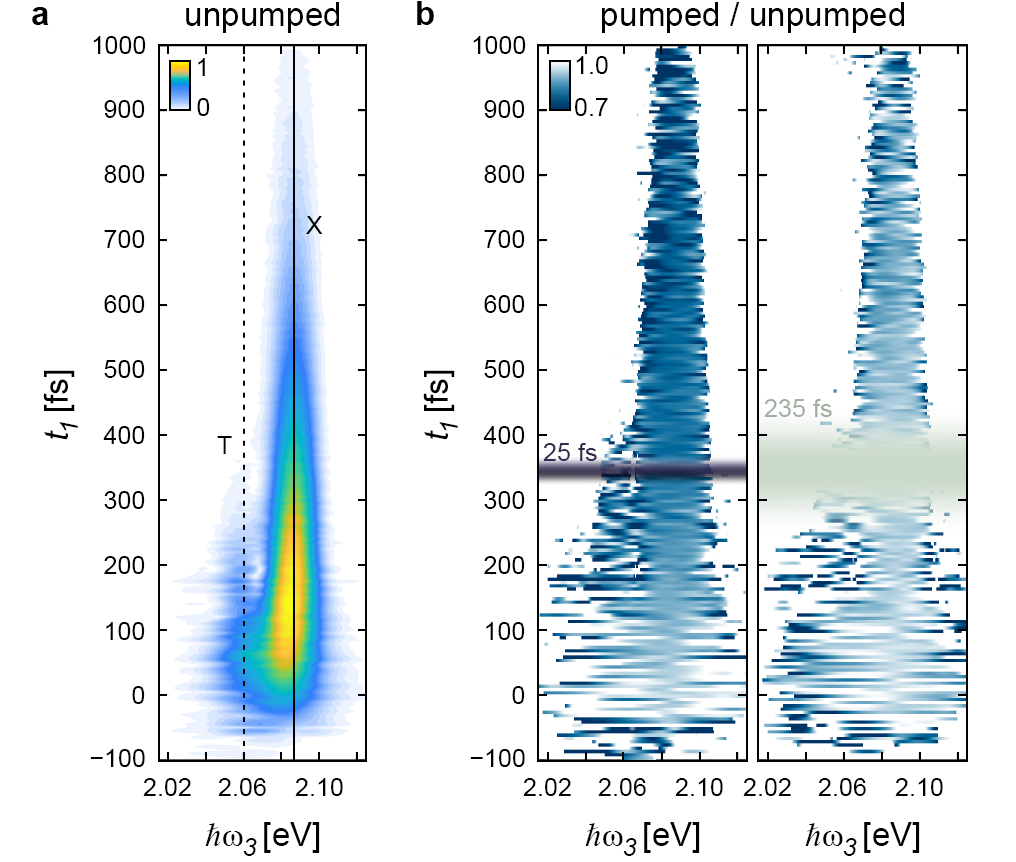}
\caption{\label{fig2} (a) Normalized 1Q-MDCS signal amplitude as a function of $t_1$ and $\hbar\omega_3$ (emission energy), in the absence of the pump. Two peaks at $\hbar\omega_3 = 2.086$ and $\hbar\omega_3 = 2.060$~eV are  attributed to the A-exciton and trion peak, respectively. (b) Ratio of pumped and unpumped 1Q-MDCS signal amplitude for a pump duration of 25~$\pm$~2 and 235~$\pm$~8~fs. The blue feature indicates that the amplitude is reduced by the pump, and this loss in macroscopic coherence is greater for a shorter pump duration (despite the same pump fluence of 205~$\mu$J/cm$^2$). This loss in macroscopic coherence is due to an increase in exciton-free carrier interactions created by two-photon absorption of the pump.}
\end{figure}

Figure~\ref{fig2}(a) shows the 1Q signal amplitude as a function of $t_1$ and $\hbar\omega_3$, in the absence of the pump. Two peaks at $\hbar\omega_3 = 2.087$ and $\hbar\omega_3 = 2.060$~eV are evident and attributed to the A-exciton and trion peak, respectively \cite{conway2023effects,muir2022interactions}. The exciton and trion energies are blue-shifted by $\sim$15 meV from the equivalent peaks in Ref.\ \cite{conway2023effects}, which arises from the difference in samples. In Ref.\ \cite{conway2023effects}, the WS$_2$ monolayer was fully encapsulated in hexagonal-Boron Nitride (hBN), while the sample used in this work has only a bottom hBN layer (see Supplementary Material for sample characterization). The unpumped signal amplitude decays as a function of $t_1$ due to decoherence of $\ket{g}\bra{X_{K^\prime}}$, with a decoherence time of $430 \pm 10$~fs, consistent with previous MDCS measurements of excitons in monolayer TMDCs \cite{conway2023effects,muir2022interactions,moody2015intrinsic,hao2016coherent}. 

Figure~\ref{fig2}(b) shows the ratio of the signal amplitude with the pump on and off, for a 25~$\pm$~2~fs and a 235~$\pm$~8~fs pump pulse arriving 350~fs before $\textbf{k}_2^{\ -}$ and $\textbf{k}_3^{\ +}$ (shaded regions in Fig.~\ref{fig2}(b)). These plots show a decrease in the signal amplitude with the introduction of the  $\sigma^+$ polarized pump, due to additional decoherence of $\ket{g}\bra{X_{K^\prime}}$. The decrease in amplitude and loss in macroscopic coherence becomes more significant for the shorter pump pulse with higher peak intensity. 

\begin{figure}
\includegraphics[width=\linewidth]{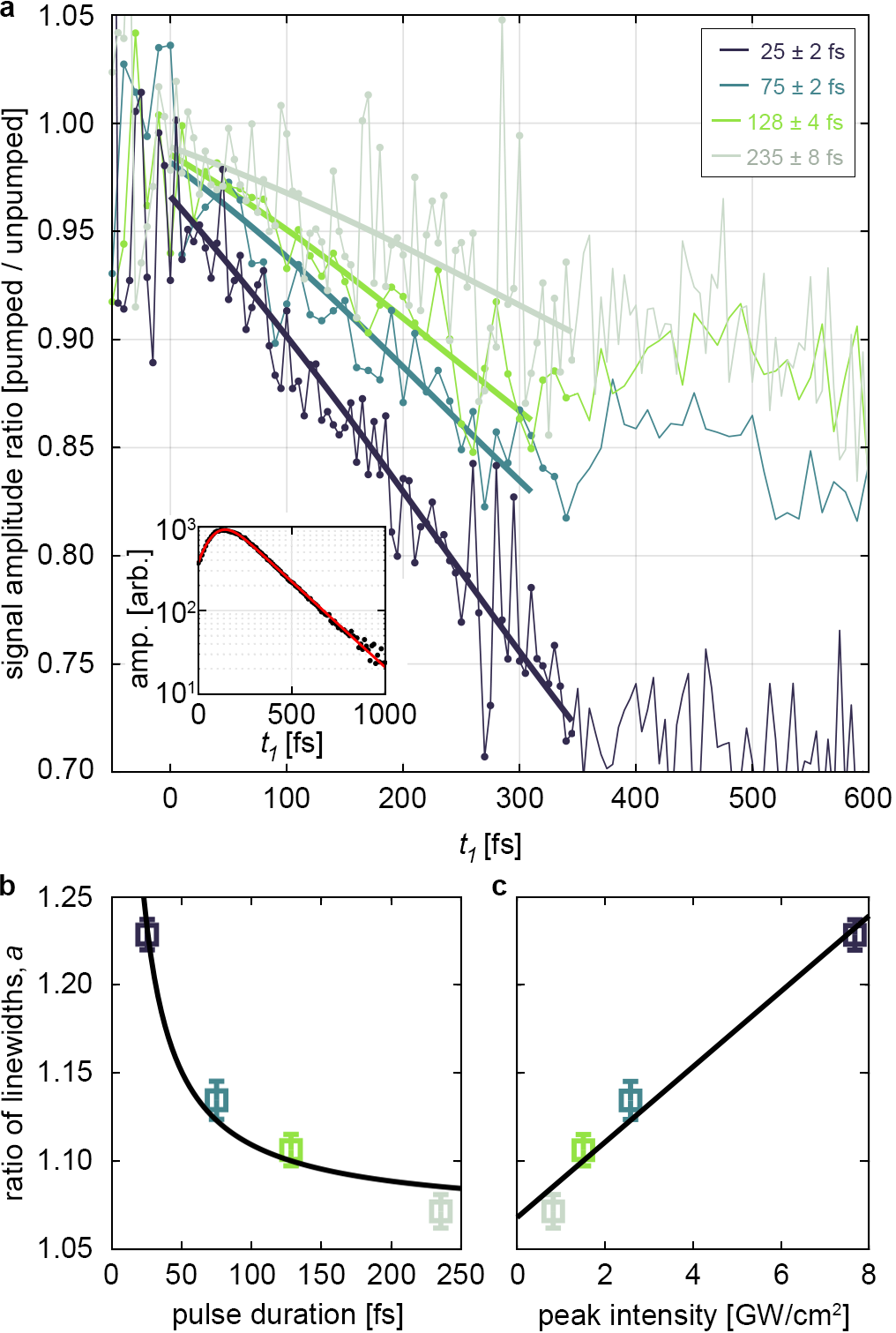}
\caption{\label{fig3} (a) Ratio of the pumped to unpumped signal amplitude (i.e.\ slices through Fig.~\ref{fig2}(b) at the exciton emission energy) fit by Eq.\ \ref{eq2} (solid lines) for each pump pulse duration. Inset: unpumped signal amplitude as a function of $t_1$  (black points) was fit with Eq.\ \ref{eq1} (red line) to determine the values for $\gamma = 1.45  \pm 0.08$~meV and $\sigma = 7.4  \pm 0.9$~meV. (b) and (c) show the values for $a$, the ratio of the homogeneous linewidths for the pump on and off, plotted as a function of pulse duration and peak intensity, respectively.}
\end{figure}

To more clearly see the dynamics of the decay and additional decoherence induced by the pump pulse, we integrate the ratio of the pumped and unpumped signal amplitudes around the exciton emission energy ($\hbar\omega_3 = 2.087  \pm 0.009$~eV), as shown in Fig.~\ref{fig3}(a). A ratio of 1 indicates the amplitude of the pumped and unpumped signal are the same, whilst ratios below 1 indicate the pump is inducing additional decoherence and decay of the signal amplitude. For $0 < t_1 < 350$~fs, the pump arrives before $\textbf{k}_1^{\ +}$ and during this period the ratio of signal amplitudes is decreasing. For $t_1 > 350$~fs the ratio remains roughly constant, indicating no further additional decoherence is present. This behaviour is consistent with extra carriers being created by the pump pulse: For $0 < t_1 < 350$~fs the $\ket{g}\bra{X_{K^\prime}}$ coherence evolves in the presence of the extra carriers created by the pump, leading to additional scattering and decoherence. For $t_1 > 350$~fs, $\textbf{k}_1^{\ +}$ arrives before the pump and the $\ket{g}\bra{X_{K^\prime}}$ coherence is not affected by the extra carriers until the pump arrives, 350~fs before $\textbf{k}_2^{\ -}$ and $\textbf{k}_3^{\ +}$. In this case, the period when there is additional scattering and decoherence is fixed, and so the ratio of pumped to unpumped signal amplitude does not change.

To quantify the additional coherence we need to take into account that the signal decay is not exponential over this time period, due to a combination of photon echo and free-induction decay. In this case the signal amplitude integrated as a function of $\hbar\omega_3$ is given by \cite{siemens2010resonance,siemens2015}:
\begin{equation}\label{eq1}
    S = \frac{e^{\frac{1}{2} \gamma (\frac{\gamma}{\sigma^2} - 4 t_1)} \sqrt{\frac{\pi}{2}} \text{erfc}[\frac{\gamma - \sigma^2 t_1}{\sqrt{2} \sigma}]}{\sigma}
\end{equation}
where $\gamma$ is the unpumped homogeneous linewidth, $\sigma$ is the inhomogeneous linewidth, and $\text{erfc}$ is the complementary error function. Inset into Fig.~\ref{fig3}(a) is the unpumped data (black points), which was fit by Eq.\ \ref{eq1} (red line) to determine the values for $\gamma = 1.45  \pm 0.08$~meV and $\sigma = 7.4  \pm 0.9$~meV.

To quantify the additional decoherence when the pump was introduced, we assumed that the decoherence rate for $0 < t_1 < 350$~fs was increased and constant for a given pulse duration. The ratio of pumped and unpumped data is then given by:
\begin{equation}\label{eq2}
       \frac{S_{\text{on}}}{S_{\text{off}}}=\frac{e^{\frac{1}{2} a\gamma (\frac{a\gamma}{\sigma^2} - 4 t_1)} \sqrt{\frac{\pi}{2}} \text{erfc}[\frac{a\gamma - \sigma^2 t_1}{\sqrt{2} \sigma}]}{e^{\frac{1}{2} \gamma (\frac{\gamma}{\sigma^2} - 4 t_1)} \sqrt{\frac{\pi}{2}} \text{erfc}[\frac{\gamma - \sigma^2 t_1}{\sqrt{2} \sigma}]}
\end{equation}
where $a$ is the ratio of the homogeneous linewidths for the pump on and off. With $a$ being the only fit parameter, Eq.\ \ref{eq2} fit very well to the data shown in Fig.~\ref{fig3}(a) to determine the increase in homogeneous linewidth, or equivalently, decoherence rate. The values for $a$ are plotted as a function of pulse duration and peak intensity in Fig.~\ref{fig3}(b) and (c), respectively. Additional decoherence due to scattering from photoexcited carriers would be expected to vary linearly with the density of carriers and thus linearly with the total fluence \cite{tollerud2017coherent}. With constant fluence, the linear increase is indicative of a process that varies quadratically with intensity, such as two-photon absorption. Briefly, this can be understood as follows: The total number of two-photon events is proportional to $\int_{-\infty}^\infty I(t)^2 dt$, where $I(t)$ is the intensity profile of the pump pulse given by $I(t) = \frac{1}{\delta\sqrt{2\pi}}\text{exp}({\frac{-t^2}{2\delta^2})}$, with the pulse duration given by the full width at half maximum, $\text{FWHM} = 2\sqrt{2\text{ln}2}~\delta$, and the peak intensity by $\frac{1}{\delta\sqrt{2\pi}}$. Rearranging this integral gives $\frac{1}{\delta\sqrt{2\pi}}\int_{-\infty}^\infty \frac{1}{\sqrt{2}}I(t) dt$. With constant pump fluence (i.e.\ $\int_{-\infty}^\infty I(t) dt$ is constant), the number of two-photon events is then linearly proportional to the peak intensity (or the inverse of the pulse duration).

The additional decoherence when the pump is on is therefore attributed to scattering with carriers excited by two-photon absorption. With the pump redshifted from the exciton transition by $\sim$260~meV, and centered at 1.82~eV, the precise identification of states available beyond 3.6~eV is challenging as it includes a multitude of high energy conduction bands in monolayer WS$_2$, and transitions beyond the $K$ points in the Brillouin zone. As a rigorous treatment is outside the scope of this work, we consider the two-photon absorption of the red-detuned pump capable of accessing continuum states which enables the creation of free carriers.

The two-photon absorption and accelerated decoherence for $0 < t_1 < 350$ observed here was not evident in the previous measurements of Ref.\ \cite{conway2023effects}. In comparison, the highest pump peak intensity used here was  $7.7  \pm 0.4$~GW/cm$^2$ for the 25~fs pump pulse duration, which is similar to the highest peak intensity reached in Ref.\ \cite{conway2023effects}, $7.1  \pm 0.5$~GW/cm$^2$, where two-photon absorption was not evident. The reason is not entirely understood, however, it is possible that the differences come from the different encapsulation environments of the samples used. The impact of top-layer hBN encapsulation on two-photon absorption in monolayer TMDCs has not been studied previously. Additionally, different degrees of aging and defects \cite{Gao2016aging,Kotsakidis2019aging,kolesnichenko2020disentangling} may influence two-photon processes. These questions are left for future work.

\subsection{Identifying bandstructure changes using the signal phase}

\begin{figure*}
\includegraphics[width=\textwidth]{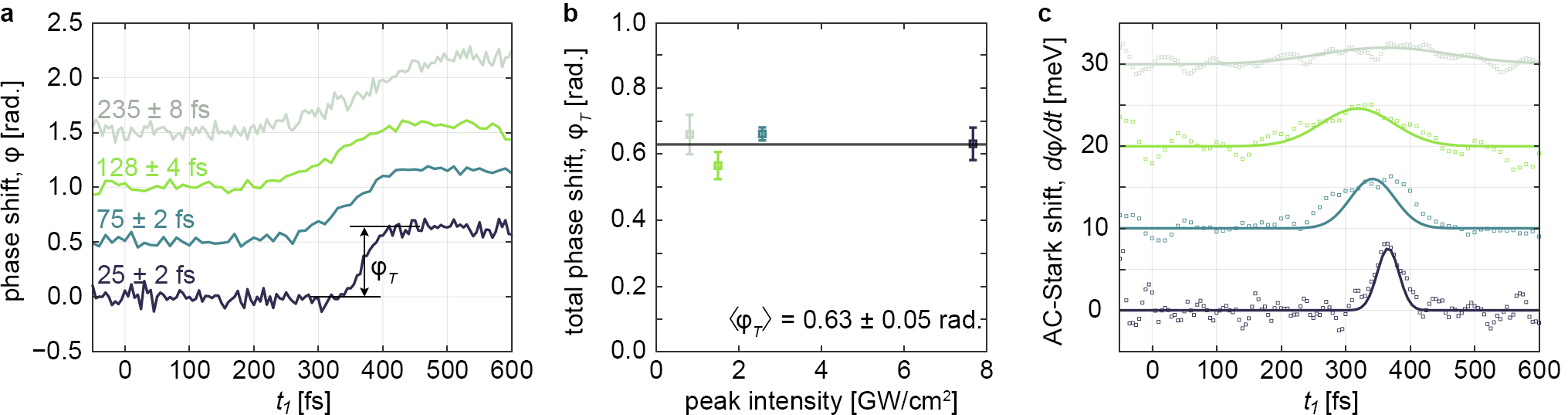}
\caption{\label{fig4}(a) Signal phase dynamics integrated around the exciton energy as a function of $t_1$, for each of the four pump pulses in Fig.~\ref{fig1}(b) (offset vertically by 0.5~rad.). (b) The magnitude of the total phase shift, $\phi_T$, is identical for each pump duration because the pump pulses have the same total fluence, consistent with the idealised Floquet formalism. We demonstrate a phase rotation of 0.63~$\pm$~0.05 rad., extracted from a difference between the phase shift in (a) before and after the pump. (c) The derivative of the phase dynamics in (a) provides the instantaneous AC-Stark shift which responds as fast as the pulse envelope. Data is smoothed by an 8-point (80~fs) moving mean and fit by a Gaussian, with FWHM extracted from a cross-correlation between the excitation pulse $\textbf{k}_1^{\ +}$ and pump pulse (39, 81, 131, and 237~fs). The fit centre and amplitude are free parameters. Each pump is offset vertically by 10~meV for clarity.}
\end{figure*}

Figure~\ref{fig4}(a) shows the evolution of the difference (pumped minus unpumped) of the measured signal phase as a function of $t_1$, averaged over the exciton energy ($\hbar\omega_3 = 2.087  \pm 0.009$~eV). For the first 200~fs there is no phase shift, until $\textbf{k}_1^{\ +}$ begins to overlap with the pump pulse. The pump, through the AC-Stark effect, blue-shifts the $K^\prime$ exciton, which causes the phase of the $\ket{g}\bra{X_{K^\prime}}$ coherence (and hence the signal) to evolve faster. The rate at which the phase shift evolves is determined by the magnitude of the blue shift, which is linearly proportional to the instantaneous pump intensity \cite{conway2023effects}. Taking the derivative of the phase shift with respect to $t_1$ gives the instantaneous blue-shift, as shown in Fig.~\ref{fig4}(c). A Gaussian fit of these data show that the measured blue-shift closely follows the cross-correlation of the $\textbf{k}_1^{\ +}$ pulse and the pump pulse, even for the shortest pump pulse. From these plots we obtain a peak AC-Stark shift ranging between 2.0 and 7.5~meV for the longest and shortest pump pulse durations, respectively.

While the peak AC-Stark shift is controlled by the peak intensity, the total accumulated phase shift is determined by integrating the blue-shift over the duration of the pump pulse. It is apparent then, that the total accumulated phase should be proportional to the pump fluence \cite{sie2015valley,kim2014ultrafast,sie2017blochsiegert,ye2017optical,cunningham2019resonant,morrow2020quantum}. In the present measurements, where the fluence was kept constant, the  accumulated phase shift should therefore be constant regardless of the pulse duration. By comparing the phase shift before and after the pump pulse ($\phi_T$ in Fig.~\ref{fig4}(a)), we obtain an averaged total phase shift of $0.63 \pm 0.05$~rad.\ which remained constant as the pulse duration (peak intensity) was changed, as shown in Fig.~\ref{fig4}(b). The magnitude of this shift is consistent with the fluence dependent phase shifts acquired in Ref.\ \cite{conway2023effects}. 

This phase response is representative of the idealised Floquet formalism, and is resolved despite the magnitude of the macroscopic coherence being reduced by more than 30\% as a result of carrier induced scattering. In fact, from the phase dynamics observed here, there is no evidence of the two-photon absorption, nor any non-adiabaticity or inconsistency with Floquet theory. Additionally, our phase response is consistent with the phase response in Ref.\ \cite{conway2023effects}, where no two-photon absorption was evident. 

\section{Conclusions}

The results presented here demonstrate the ability of MDCS to quantify changes to the bandstructure of monolayer WS$_2$ induced by Floquet engineering, even in the presence of significant broadening. This was achieved by following the evolution of the average phase of the macroscopic coherence excited by the MDCS pulses. Furthermore, we were able to quantify the decoherence caused by carriers excited by two-photon absorption of the Floquet pump pulses, through the evolution of the signal amplitude. Being able to follow the evolution of the bandstructure, while at the same time quantifying the effect of excess carriers is an important step towards understanding the role of excited carriers in heating and driving nonadiabatic effects in Floquet engineering of solid state systems. Previous work has shown that scattering can prevent the formation of Floquet-Bloch bands when the scattering time is faster than the period of the driving field \cite{aeschlimann2021survival}. Even with the excess carriers, we are far from this regime, with driving period of 2.3~fs and scattering time in the hundreds of fs. Further, based on measurements of the exciton-exciton interaction strength in monolayer TMDCs \cite{moody2015intrinsic}, we expect that even with carrier densities beyond $10^{13}$~cm$^{-2}$ the scattering time will remain longer than the drive period, which enhances the prospect that these excess carriers can be used to help shape the new potential energy environment and resultant band structures \cite{seetharam2015controlled,rudner2020band}. 
Finally, these measurements indicate that is should be possible to measure the effects of a Floquet-topological phase transition in monolayer TMDCs, even though a high density of carriers, excited by the above bandgap drive, are inevitable.


\section{Acknowledgements}

This work was funded by the Australian Research Council Centre of Excellence in Future Low-Energy Electronics Technologies (CE170100039). T.H.Y.V., M.S.F., and M.T.E.\ acknowledge funding support from DP200101345. This work was performed in part at the Melbourne Centre for Nanofabrication (MCN), the Victorian Node of the Australian National Fabrication Facility (ANFF). K.W.\ and T.T.\ acknowledge support from JSPS KAKENHI (Grant Numbers 21H05233 and 23H02052) and World Premier International Research Center Initiative (WPI), MEXT, Japan.


\bibliography{references}

\end{document}